\documentclass{article} 
\usepackage{colm2024_conference}
\usepackage{microtype}
\usepackage{hyperref}
\usepackage{url}
\usepackage{booktabs}
\usepackage{amsmath,amssymb,amsfonts}
\usepackage{algorithmic}
\usepackage{graphicx}
\usepackage{textcomp}
\usepackage{xcolor}
\usepackage{multicol}
\usepackage{multirow}
\usepackage{tcolorbox}
\usepackage{lipsum}
\usepackage{newfloat}
\usepackage{listings}
\usepackage{tabularx}
\usepackage{tcolorbox}
\usepackage{xcolor}
\usepackage{hyperref}

\definecolor{darkblue}{rgb}{0, 0, 0.5}
\hypersetup{colorlinks=true, citecolor=darkblue, linkcolor=darkblue, urlcolor=darkblue}

\newtcolorbox{mybox}[2][]{
  colback=gray!10!white,
  colframe=black!75!black,
  fonttitle=\bfseries,
  coltitle=white,
  rounded corners,
  boxrule=0.5mm,
  title=#2,#1,
}

\newtcolorbox{promptbox}{
  colback=gray!5!white, 
  colframe=gray!75!black, 
  rounded corners, 
  boxrule=0.3pt, 
  coltitle=black,
}

\def\BibTeX{{\rm B\kern-.05em{\sc i\kern-.025em b}\kern-.08em
    T\kern-.1667em\lower.7ex\hbox{E}\kern-.125emX}}

\title{Advancing TTP Analysis: Harnessing the Power of Large Language Models with Retrieval Augmented Generation}

\colmfinalcopy
\author{Reza Fayyazi, Rozhina Taghdimi \& Shanchieh Jay Yang \\
Department of Electrical \& Computer Engineering\\
Rochester Institute of Technology\\
Rochester, NY, USA \\
\texttt{rf1679@rit.edu, rt3271@rit.edu, jay.yang@rit.edu} \\
}

\colmfinalcopy 
\begin{document}

\maketitle

\begin{abstract}
Tactics, Techniques, and Procedures (TTPs) outline the methods attackers use to exploit vulnerabilities. The interpretation of TTPs in the MITRE ATT\&CK framework can be challenging for cybersecurity practitioners due to presumed expertise and complex dependencies. Meanwhile, advancements with Large Language Models (LLMs) have led to recent surge in studies exploring its uses in cybersecurity operations. It is, however, unclear how LLMs can be used in an efficient and proper way to provide accurate responses for critical domains such as cybersecurity. This leads us to investigate how to better use two types of LLMs: small-scale encoder-only (e.g., RoBERTa) and larger decoder-only (e.g., GPT-3.5) LLMs to comprehend and summarize TTPs with the intended purposes (i.e., tactics) of a cyberattack procedure. This work studies and compares the uses of supervised fine-tuning (SFT) of encoder-only LLMs vs. Retrieval Augmented Generation (RAG) for decoder-only LLMs (without fine-tuning). Both SFT and RAG techniques presumably enhance the LLMs with  relevant contexts for each cyberattack procedure. Our studies show decoder-only LLMs with RAG achieves better performance than encoder-only models with SFT, particularly when directly relevant context is extracted by RAG. The decoder-only results could suffer low `Precision' while achieving high `Recall'. Our findings further highlight a counter-intuitive observation that more generic prompts tend to yield better predictions of cyberattack tactics than those that are more specifically tailored. \footnote{Code available at: https://github.com/RezzFayyazi/TTP-LLM}

\end{abstract}

\section{Introduction}
The evolving changes of cybersecurity vulnerabilities presents a significant challenge for incident and threat analysis. 
Security analysts utilize Cyber Threat Intelligence (CTI) to gather, analyze, and interpret the Tactics, Techniques, and Procedures (TTPs) of adversaries. The process of understanding the various cyberattack procedures (which is constantly updated in reference models like MITRE ATT\&CK \cite{MITRE}), requires extensive expertise and effort. 
The ATT\&CK framework utilizes a 3/4-tier Tactic-Technique/Sub-technique-Procedure hierarchy to explain the \textbf{how} and \textbf{why} behind attackers' exploitation of vulnerabilities. 
However, the complexity and potential ambiguity of these descriptions can make them challenging to interpret, leading to differing conclusions among analysts from the same description. For example, consider the following attack procedure description:

\vspace*{-10pt}

\noindent 
\begin{quote}
{\it 
\small Threat Group-3390 has performed DLL search order hijacking to execute their payload.
}
\end{quote}

\vspace*{-2pt}

\noindent The ATT\&CK framework associates three tactics with this procedure, namely, `Privilege Escalation', `Defense Evasion', and `Persistence'. It is not obvious that the DLL Search Order Hijacking technique will lead to any of the three, not to mention all three adversarial intents. In fact, some might argue that this description is more aligned with the `Execution' tactic, as the adversary tries to execute its malicious payload. 

Therefore, in this work, we delve into the potential of Large Language Models (LLMs) to interpret and map procedure descriptions into specific ATT\&CK tactics. Note that we recognize the inherent imperfection of ATT\&CK TTP mapping, and this, in fact, motivated us to study the capabilities and limitations of LLMs under such real-world constraints. 
The evolution of LLMs, such as the OpenAI's GPT-3.5 model \cite{OpenAI}, has led to substantial progress in NLP tasks \cite{min2023recent, Zhao2023}. This progress is mainly due to the LLM's enhanced semantic understanding and scalability.  There have also been works that process TTPs by leveraging LLMs \cite{Rani2023, Alves2022, Orbinato2022}. However, none of these works compared the use of encoder-only LLMs (e.g., RoBERTa) and decoder-only LLMs (e.g., GPT-3.5) while leveraging Retrieval Augmented Generation (RAG) techniques to analyze the descriptions of attack procedures. 

The encoder-only LLMs (e-LLM) excel in tasks like sequence classification by processing input data to generate contextual representations, while decoder-only models (d-LLM) are adept at generating coherent, contextually relevant text, making them ideal for language generation tasks. One major problem of d-LLMs is the tendency to hallucinate \cite{huang2023survey, rawte2023troubling}. This issue involves the model's ability to produce responses that are misleading or completely fabricated, yet presented confidently. To address this issue, RAG techniques are proposed to extract the most relevant documents and feed those as additional information into part of the LLM prompts. It is used to shift the model's attention to relevant information by first retrieving texts that are likely to contain relevant information to generate a more validated response. Note that we are not fine-tuning the d-LLMs, as such a process requires significant computational resources. Instead, we consider RAG to be a potentially efficient technique to provide d-LLMs with cybersecurity context.

Recognizing the potentially different uses and benefits for e-LLMs and d-LLMs for cybersecurity operations, this study sets out to compare the supervised fine-tuning (SFT) of e-LLMs and RAG with d-LLMs to interpret TTPs in their effectiveness to interpret cyberattack procedures (assessed by predicting ATT\&CK tactics). Specifically, we examine four sets of approaches: 1) direct use of the d-LLMs (GPT-3.5), 2) SFT of e-LLMs (RoBERTa \cite{Liu2019} and SecureBERT \cite{Aghaei2023}) with ATT\&CK descriptions, and 3) the use of RAG on d-LLMs and 4) analysis of d-LLM capabilities with more specific vs. more generic prompts to predict cyberattack tactics. 

To the best of our knowledge, this is the first formal study to compare these approaches to interpret cyberattack procedure descriptions. We aim to derive `practical' approaches that are reasonable for cybersecurity practitioners to adopt. 
By experimenting using the TTP descriptions extracted from MITRE ATT\&CK (release v14.1 as of Oct. 31, 2023), we present the benefits and limitations of the two LLM architectures and suggest future research directions to enhance d-LLM's use in cyber-ops.  Our contributions includes:

\begin{itemize}
    \item We analyze and compare the performance in SFT of smaller e-LLMs vs. RAG-enhanced of larger d-LLMs for interpreting TTPs.
    \item We demonstrate d-LLMs achieve a superior performance in interpreting TTPs when directly relevant information is extracted via RAG.
    \item We show that while d-LLMs maintain strong 'Recall', they lack 'Precision', and we elaborate the need for new advances in increasing the `precision' in d-LLMs without compromising `recall'. 
    \item We present preliminary results showing that the use of generic prompts seemingly provide more informative and accurate responses than the use of specific prompts.
\end{itemize}


\section{Related Works}
\label{sec:related}

\subsection{Large Language Models}

The advancements in Large Language Models are primarily due to the adoption of the Transformer architecture \cite{vaswani2017attention}. Central to this architecture are the encoder and decoder components, which is utilized distinctively in various LLMs.
 
\textbf{Encoder-Only}: One of the pioneering introductions is BERT \cite{Devlin2018}, a pre-trained bidirectional e-LLM based on Transformers. It leverages the Transformer encoder to achieve a deep understanding of language context and nuances. 
Moreover, RoBERTa \cite{Liu2019}, extended BERT's capabilities with optimized hyperparameters and enhanced training with larger mini-batches and longer sequences. 
Another notable e-LLM is SecureBERT \cite{Aghaei2023}, which is based on RoBERTa but specifically fine-tuned with cybersecurity data.

\textbf{Decoder-Only Models:} The landscape of LLMs has been further transformed by d-LLMs, characterized by their extensive training datasets and significant computational resources. OpenAI's GPT-3.5 \cite{OpenAI} and Meta's LLAMA-2 \cite{touvron2023llama} are leading examples in this category, both with proficiency in generating human-like text and recognizing intricate patterns across various fields. 
The effectiveness and versatility of these models in handling diverse and complex tasks have been extensively documented in the literature, as described in \cite{min2023recent, Zhao2023}. However, there remains a question on how capable these models are in interpreting TTPs.

\subsection{LLM for TTP interpretations}

In the context of processing TTP descriptions, several works have been proposed to use NLP techniques. \cite{Sauerwein2022, Husari2017, Kim2022}. More recently, the following works have utilized e-LLMs (e.g., BERT) for classifying TTPs. 
Alves et al. \cite{Alves2022} tested various BERT model variants with different hyperparameters to find the the best model for TTP classification. 
Rani et al. \cite{Rani2023} developed TTPHunter, a tool that uses BERT and RoBERTa models with a linear classifier for extracting TTPs from unstructured text.
Orbinato et al. \cite{Orbinato2022} conducted an experimental study to compare traditional ML and DL models (e.g., SVM vs. SecureBERT) for mapping TTPs from CTI texts. The authors found SecureBERT to be the most effective, however, they did not provide a comparison between SecureBERT and its foundational RoBERTa model.

Our preliminary study \cite{fayyazi2023uses} compared e-LLMs vs. the direct use of d-LLMs (i.e., prompt engineering) to process and classify higher-level tactic and technique descriptions (but not procedures). In that study, we fine-tuned the e-LLM classifier by drawing from the pool of tactic and technique descriptions, and the d-LLM approach did not include with RAG. This paper significantly expands our prior work's findings to treat harder-to-interpret attack procedure descriptions. It adopts a more realistic setting where procedure descriptions are not used for fine-tuning and incorporates RAG techniques.

\subsection{Retrieval Augmented Generation (RAG)}

Despite the proficiency of d-LLMs in instruction-following tasks (e.g., question-answering), these models have shown to be prone to hallucination in their generated responses \cite{huang2023survey, rawte2023troubling}. Their reliance on pre-trained knowledge (i.e., constrained by the training data's temporal scope) poses limitations. This limitation is critical in fields requiring high precision, such as cybersecurity. 
Therefore, to adapt to new information and to provide factual knowledge, RAG was introduced \cite{borgeaud2022improving}.

RAG typically utilizes a vector database to store up-to-date information. Upon receiving a query, RAG selects relevant data from this database to enrich the LLM's prompt, thereby ensuring responses are both current and contextually grounded without constant fine-tuning. 
There have been works that used RAG in NLP tasks \cite{borgeaud2022improving, mallen2023not, ram2023context, al2023transforming}. In this work, we propose the use of RAG to provide more information about an attack procedure to the LLM and shift the LLM's attention towards the text retrieved through RAG. 

\section{Methodology \& Experimental Design}
\label{sec:method}
We design this study to compare: 1) supervised fine-tuning (SFT) of e-LLMs with labeled ATT\&CK technique/sub-technique descriptions, 2) direct use of d-LLMs (i.e., pre-trained knowledge), 3) use of RAG for d-LLMs with relevant URLs retrieved by finding the most similar attack procedures, 4) analysis of d-LLM capabilities in interpreting attack procedures using a specific vs. generic prompt. 
This experimental design enables us to assess how well LLMs interpret cyberattack procedure descriptions and map them to the corresponding tactic(s). This is a challenging process due to multiple reasons: 1) there are no publicly available datasets that map procedure descriptions into ATT\&CK tactic(s), 2) each procedure can map to multiple ATT\&CK tactics that LLMs need to consider, 3) a well-structured prompt is needed for LLMs to generate desired outputs, and 4) RAG needs to retrieve the most relevant information. We outline our methodology and experimental design to tackle these challenges.

\subsection{Datasets}

\begin{figure}[htb]
\centering
\includegraphics[scale=0.30]{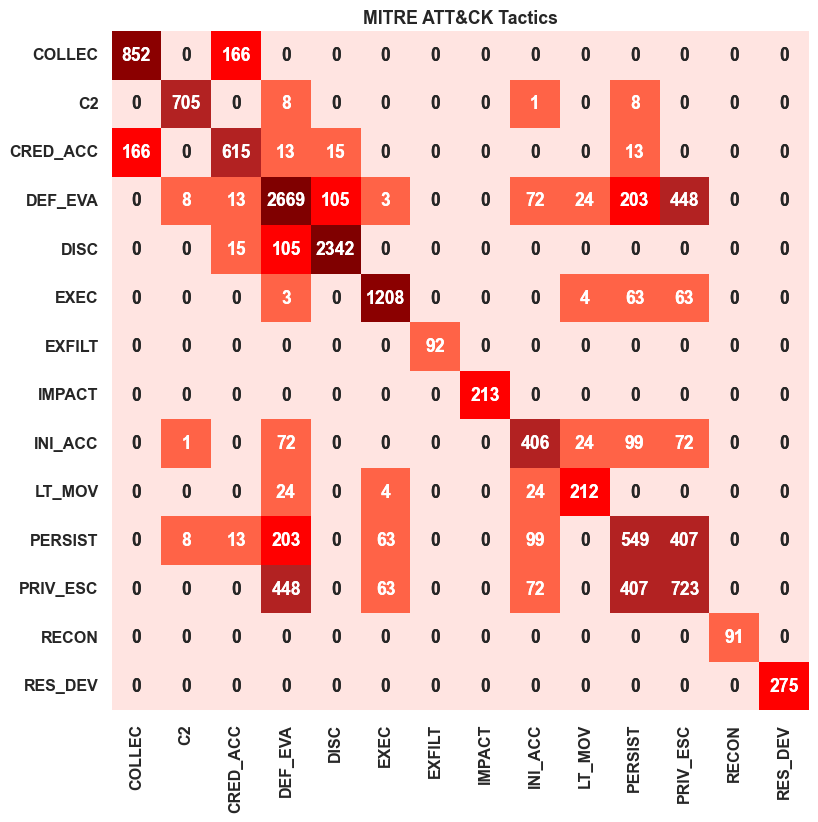}
\caption{The pair-wise overlap of tactics on ATT\&CK procedure descriptions.}
\label{fig:overlaps}
\end{figure}

To conduct our experiments, we gathered data from the MITRE ATT\&CK framework. We chose ATT\&CK due to its broad adoption in industry SIEM tools and associated detailed cybersecurity descriptions. We curated the descriptions of enterprise tactics, techniques, and sub-techniques along with the mapping to their corresponding tactic(s) for fine-tuning the e-LLMs. We do not fine-tune the targeted d-LLMs (GPT-3.5); instead, we use RAG for d-LLMs to compared to the fine-tuned e-LLM. In total, we obtained 639 descriptions from the ATT\&CK framework for fine-tuning e-LLMs. Some descriptions have 2,3, or even 4 different tactics. Note that the small volume of 639 labeled ATT\&CK descriptions reflects the real-world scenario where only limited labeled data would exist for targeted fine-tuning. 
To test the performance of both e-LLMs and d-LLMs, we crawled all the enterprise procedure examples that describe an attack methodology from the ATT\&CK framework. We removed the procedure descriptions that contain any of the 14 MITRE ATT\&CK tactics within the descriptions to prevent potential biases or shortcuts for the LLMs. In total, we gathered 9,532 procedure descriptions along with their corresponding URLs (for RAG). Note that the procedure descriptions are not used for SFT of e-LLMs. Figure \ref{fig:overlaps} shows the pair-wise overlap across tactics for attack procedure descriptions. As can be seen, `Persistence', `Privilege-Escalation' and `Defense-Evasion' tactics have higher overlaps with the others. We will share the curated data along with our source code as artifacts for the research community. 

\subsection{Supervised Fine-Tuning of Encoder-Only LLMs}

\begin{figure*}[hbt]
\centering
\includegraphics[scale=0.30]{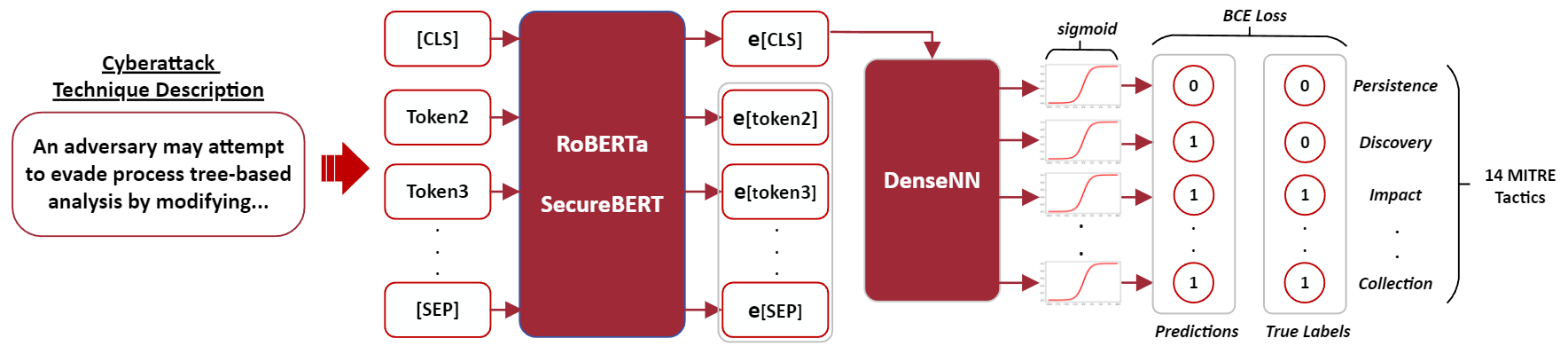}
\caption{The process of supervised multi-label fine-tuning of e-LLMs with high-level ATT\&CK descriptions (excluding `procedures') -- adopted from \cite{fayyazi2023uses}.}
\label{fig:supervisedtraining}
\end{figure*}

\vspace*{-10pt}

To conduct supervised fine-tuning with e-LLMs, we selected RoBERTa-base \cite{Liu2019} and SecureBERT \cite{Aghaei2023}. We chose these models because of their demonstrated superiority over traditional ML and DL techniques in processing TTPs \cite{Rani2023, Alves2022, Orbinato2022}. The SecureBERT model has already fine-tuned using an extensive cybersecurity text corpus. 
Our goal is to assess these LLMs' efficacy in distilling crucial information through fine-tuning with the most relevant ATT\&CK Tactics, Techniques, and Sub-techniques descriptions (but not procedures). 
Figure \ref{fig:supervisedtraining} shows the overall process of fine-tuning both RoBERTa and SecureBERT models. 
First, we tokenize the ATT\&CK descriptions for compatibility with both RoBERTa and SecureBERT models. Second, we add a classification layer with 14 nodes with each corresponding to an ATT\&CK tactic. To address the challenges of multi-label classification, we employ a Sigmoid activation function within this layer. This function is crucial as it calculates the independent probability of each tactic for a given description. This enables the models to effectively discern and assign multiple labels where appropriate.  

Table \ref{tab:hyperparameters} shows the hyperparameters selected to fine-tune the models. We employed binary cross-entropy as the loss function (due to the binary nature to predict each tactic), 16 for the batch size, 30 epochs, and 5e-5 for learning\_rate as the best-performing values. Model outputs are 14-dimensional binary vectors, with each element representing an ATT\&CK tactic, with a Sigmoid output over 0.5 indicating a prediction for that specific tactic.

\vspace*{-8pt}

\begin{table}[htbp]
\small
\centering

\caption{The hyperparameters tested and selected for SFT of RoBERTa and SecureBERT.}
\label{tab:hyperparameters}
\begin{tabular}{c|c}
\textbf{Hyperparameter} & \textbf{Value} \\
\hline
Optimizer & AdamW \\
Loss Func. & BCE \\
Activation Func. & Sigmoid \\
Batch Size & [8, \textbf{16}] \\
Epoch & [25, \textbf{30}, 35] \\
Learning rate & [1e-5, 2e-5, \textbf{5e-5}, 7e-7] \\
\end{tabular}
\end{table}

\vspace*{-8pt}

\subsection{Decoder-Only LLMs with and w/o RAG}

We consider OpenAI's GPT-3.5-turbo-1106 model which has a context window of 16K tokens, as our d-LLM. It is worth noting that in OpenAI models, their inherent design are non-deterministic. This means that identical inputs can result in different outputs. Therefore, to manage this aspect and have consistency, we set the temperature parameter to `0' and a seed number (1106) to ensure reproducibility. Our baseline use of GPT-3.5-Turbo will be without RAG.  We engineered a simple and clean prompt without overly influencing the attention of the LLM to the prompt other than the procedure description, while attempting to avoid hallucination based on best practices reported in the community. Below is the engineered prompt used for the d-LLM baseline:

\begin{promptbox}
\small
{\it You are a cybersecurity expert.

Knowing that $<<${procedure}$>>$, what MITRE ATT\&CK tactics will a cyber adversary achieve with this technique?

Please only respond with the MITRE ATT\&CK tactics you are certain about.
}
\end{promptbox}

\begin{figure*}[t]
\centering
\includegraphics[scale=0.30]{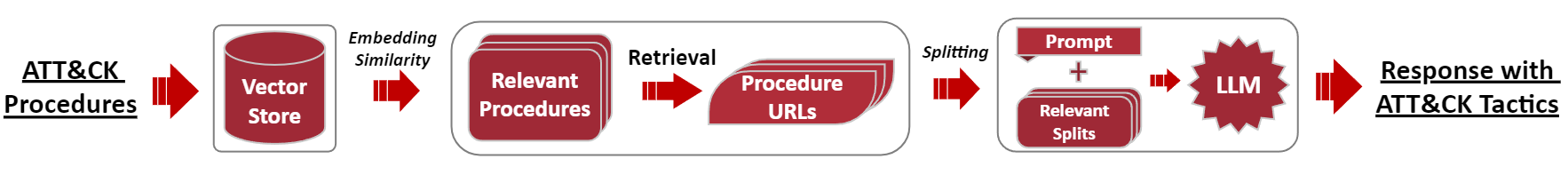}
\caption{The overall process of RAG with relevant ATT\&CK procedures of d-LLMs.}
\label{fig:retrieval}
\end{figure*}

We design our d-LLM with RAG with an attempt to mimic a real-world scenario where the procedure in question does not exist and cannot be searched directly to find the corresponding tactic(s). Figure \ref{fig:retrieval} shows the overall process of using RAG with d-LLMs. We utilize FAISS \cite{Faiss} to find top-3 similar procedures in our dataset for each procedure tested. These 3 similar procedures will provide up to 3 URLs where RAG will use to retrieve chunks of text serving as contexts for the prompt. We elected to use 3 chunks of 8,000 characters with chunk\_overlap of 500, 
and store them in a Vector Store. Then, we use the OpenAI embeddings for both the question and the stored chunks, and we retrieve the top-3 most relevant chunks based on the question and insert them as `Relevant Context' into the prompt. The prompt template used for GPT-3.5-Turbo is as follows:  

\begin{promptbox}
\small
{\it You are a cybersecurity expert. Consider the relevant context provided below and answer the question.

Relevant Context: \{context\}

Question: Knowing that $<<${procedure}$>>$, what MITRE ATT\&CK tactics will a cyber adversary achieve with this technique?

Please only respond with the MITRE ATT\&CK tactics you are certain about.
}

\end{promptbox}

\noindent Note that in this case, some of the similar procedures can have the same URL as the procedure in question. This could lead to GPT-3.5-Turbo finding direct answer through RAG. However, as we will demonstrate in the results, the performance is good but still not perfect even if the exact URL is among the ones found. We also note that the ATT\&CK's techniques/sub-techniques URLs actually contain the tactic names. 

\vspace*{-2pt}

To establish an optimal performance benchmark for RAG, we conducted tests in an idealized scenario where RAG is supplied with only the exact URL containing the target procedure. This enforces the selection of top-3 8,000 character chunks to come from the most relevant URL. We note that some URLs have less text and others have many. This variation may also influence the performance of RAG. Nevertheless, as we will show in the results, this ideal case achieves excellent upper bound performance.

\vspace*{-4pt}

\section{Results \& Discussion}
\label{sec:results}

\vspace*{-8pt}

We first evaluate which e-LLM performs better. Then, we compare the d-LLMs with and without RAG. Next, we compare the capabilities and limitations of the selected e-LLM and d-LLMs with RAG. Finally, we provide analysis of some specific cases and discuss the effect of more flexible prompts on d-LLMs' capabilities.  
This work considers the `Samples Average F1' score. F1 score captures the harmonic mean of Precision and Recall. We consider `Recall' to assess LLM's capability to interpret procedures within a broader context, and `Precision' to assess its capability to map them to the exact tactics. The Samples Average F1 Score calculates the F1 score for each instance and then averages these scores. The reason for using the Samples Average, as opposed to Micro, Macro and Weighted average scores, is that it reduces the impact of an imbalanced number of procedures across tactics or over-emphasis of procedures that have multiple labels.

\subsection{Evaluation of SFT of Encoder-Only LLMs}
We fine-tune the e-LLMs (RoBERTa-SFT and SecureBERT-SFT) using the curated MITRE ATT\&CK enterprise Tactics, Techniques, and Sub-techniques descriptions (mapped to AT\&CK's tactics). We should emphasize that we do not fine-tune these models with procedure descriptions, as we want to evaluate the `interpretation' capability of these LLMs (i.e., by not giving these models the explicit \textit{procedure, tactic} mappings). For the results, we obtained \textbf{0.54} and \textbf{0.41} as Samples Average F1 score for SecureBERT and RoBERTa, respectively. SecureBERT exhibited stronger performance compared to the RoBERTa model, and is chosen to compare to the d-LLMs. 

\vspace*{-8pt}


 



\subsection{Prompt-Only vs. RAG of GPT-3.5-Turbo}


Table \ref{table:decoderonly_performance} shows the performance of the aforementioned test cases for all 9,532 ATT\&CK procedure descriptions separated by tactics they are associated with according to ATT\&CK. Note that since some procedures are associated with multiple tactics, the sum of all supports (10,952) exceeds the number of tested procedures (9,532). As the results suggest, using only the prompt (i.e., relying on GPT-3.5's pre-trained knowledge) results in a poor performance in interpreting the ATT\&CK tactics with `0.60' Samples Avg. F1 (but higher than SecureBERT-SFT). In 4 out of 14 tactics, it achieved a F1 score higher than `0.50', but got `0.00' for Resource Development. In contrast, when using RAG by providing the exact URL (the ideal case), the model's performance for Samples Avg. F1 significantly improved to `0.95', with 11 out of 14 tactics got F1 higher than `0.90'. 

This indicates that by providing the model with the accurate information (chunks of data that contain the answer), it achieves significantly better performance without the need for fine-tuning. However, in a more realistic scenario (RAG with top-3 similar procedure), where the exact piece of information of a `new' attack procedure are not easily accessible, the performance falls down to `0.68'  Samples Avg. F1 Score (while still superior than prompt-only). This implies that one should have limited expectation for LLM to interpret new, likely unseen attack procedures even with RAG.

It is worth noting that the d-LLM responses include more than just the tactics' names. We applied a keyword search of the ATT\&CK tactics for this study, and ignore the potential analogous terms (due to their generative capability). Our examination of d-LLM responses shows very few such cases, and we estimate a potential small deviation ($\sim$1\%) in our performance measure if we were to apply a thorough extraction of analogous terms.

\vspace*{-8pt}

\begin{table*}[hbt]

\begin{center}
\small
\caption{Per-tactic performance of Prompt-Only vs. RAG w/ Reference URL vs. RAG w/ Relevant Procedures with the GPT-3.5 model for all ATT\&CK procedure descriptions.}

\setlength{\tabcolsep}{2pt} 
\renewcommand{\arraystretch}{1.00}

\begin{tabular}{@{}c !{\vrule width 1pt} c !{\vrule width 1pt} c !{\vrule width 1pt} c !{\vrule width 1pt} c@{}} 
  \textbf{GPT-3.5-Turbo} & \textbf{Prompt-Only} & \textbf{RAG w/ Ref. URL} & \textbf{RAG w/ Relevant Procedures} \\ 
\hline

\textbf{Tactics} & \textbf{F1 Score} & \textbf{F1 Score} & \textbf{F1 Score} & \textbf{Support} \\ 
\hline
 Collection  & 0.44  & 0.84 & 0.53 & 852 \\
 C2 & 0.52  & 0.96 & 0.59 & 705  \\
 Credential Access & 0.59  & 0.92 & 0.69 & 615 \\
 Defense Evasion & 0.66  & 0.96 & 0.76 & 2669 \\
 Discovery & 0.86 & 0.98 & 0.85 & 2342\\
 Execution & 0.48 & 0.90 & 0.55 & 1208 \\
 Exfiltration & 0.20  & 0.65 & 0.31 & 92 \\
 Impact & 0.44 & 0.84 & 0.54 & 213 \\
 Initial Access & 0.42  & 0.91 & 0.46 & 406  \\
 Lateral Movement & 0.45 & 0.90 & 0.44 & 212 \\
 Persistence & 0.40 & 0.92 & 0.49 & 549 \\
 Privilege Escalation & 0.41  & 0.92 & 0.60 & 723  \\
 Reconnaissance & 0.21 & 0.99 & 0.42 & 91\\
 Res. Development & 0.00  & 0.93 & 0.50 & 275  \\
 \hline
  \textbf{Samples Avg. F1} & \textbf{0.60}  & \textbf{0.95} & \textbf{0.68} & \textbf{10952}  \\
 
\end{tabular}
\label{table:decoderonly_performance}
\end{center}
\end{table*}

\vspace*{-8pt}


\subsection{Decoder-Only w/ RAG vs. SFT of Encoder-Only}

We further compare the SFT of e-LLM (SecureBERT) vs. d-LLM with RAG models by analyzing their performance based on Precision, Recall, and F1 scores. For d-LLM with RAG, we separate the 9,532 procedures into two sub-groups: 1) at least one top-3 URL match to the exact URL of the procedure in question, and 2) no top-3 URL matches to the exact URL. This separation gives a detailed insights on how RAG performs with and without the exact URL, and to provide a comparison on the `main' differences between these e-LLMs and d-LLMs. 

Table \ref{table:encoderdecodercomparison} shows their performances. The e-LLMs perform a lot better in `Precision' than `Recall'. This means that SecureBERT (the e-LLM) tries to provide precise responses. However, as shown in GPT-3.5 (the d-LLM) results, they perform significantly better for `Recall' than `Precision'. This can potentially indicate that hallucinations in LLMs typically arise during the decoding phase, where the model tries to generate information not strictly present/implied in the input. This is a major factor that can reduce precision. Therefore, this indicates a need for improving `Precision' in d-LLMs without compromising `Recall'.

\begin{table*}[hbt]
\small
\begin{center}
\caption{Per-tactic performance of e-LLM and d-LLM for ATT\&CK procedure descriptions. For GPT-3.5, we compare cases where the source and retrieved procedures share the same URL (4115/9532 procedures) and where they do not (5417/9532 procedure). 
}
\begin{scriptsize}
\setlength{\tabcolsep}{1.5pt} 
\renewcommand{\arraystretch}{1.00} 
\begin{tabular}{c !{\vrule width 1pt} c | c | c | c !{\vrule width 1pt} c | c | c | c !{\vrule width 1pt} c | c | c | c } 

  \textbf{Model:} & \multicolumn{8}{c}{\textbf{GPT-3.5}} & \multicolumn{3}{c}{\textbf{SecureBERT}} & \\ 
\hline
  \textbf{Data:} & \multicolumn{4}{c}{\textbf{w/ Matched URLs}} & \multicolumn{4}{c}{\textbf{w/o Matched URLs}} & \multicolumn{3}{c}{\textbf{SFT w/ Tactic \& Techniques}} & \\ 
\hline
\textbf{Tactics} & \textbf{Precision} & \textbf{Recall} & \textbf{F1} & \textbf{Support} & \textbf{Precision} & \textbf{Recall} & \textbf{F1} & \textbf{Support} & \textbf{Precision} & \textbf{Recall} &  \textbf{F1} & \textbf{Support} \\ 
\hline 
Collection & 0.58 & 0.96 & 0.72 & 354 & 0.31 & 0.67 &  0.42 & 498 & 0.56 & 0.56 & 0.56 & 852 \\
C2 & 0.73 & 0.96  & 0.83 & 314 & 0.33 & 0.55  & 0.41 & 391 & 0.74 & 0.26 & 0.39 & 705  \\
Credential Access & 0.79 & 0.99  & 0.88 & 268 & 0.46 & 0.71  & 0.56 &  347 & 0.69 & 0.71 & 0.70 & 615 \\
Defense Evasion & 0.87 & 0.97  & 0.92 & 1162 & 0.59 & 0.71  & 0.65 & 1507 & 0.80 & 0.60 & 0.68 & 2669 \\
Discovery & 0.95 & 0.97  & 0.96 & 1111 & 0.73 & 0.76  & 0.74 & 1231 & 0.95 & 0.69 & 0.80 & 2342\\
Execution & 0.71 & 0.89  & 0.79 & 548 & 0.31 & 0.54 & 0.39 & 660 & 0.63 & 0.48 & 0.54 & 1208 \\
Exfiltration & 0.30 & 0.96 & 0.46 & 28 & 0.16 & 0.67  & 0.26 & 64 & 0.41 & 0.42 & 0.42 & 92 \\
Impact & 0.66 & 0.96  & 0.78 & 118 & 0.23 & 0.59 & 0.33 & 95 & 0.83 & 0.23 & 0.35 & 213 \\
Initial Access & 0.57 & 0.96  & 0.71 & 156 & 0.24 & 0.59  & 0.34 & 250 & 0.62 & 0.27 & 0.37 &  406  \\
Lateral Movement & 0.51 & 0.97  & 0.67 & 71 & 0.24 & 0.48  & 0.32 & 141 & 0.46 & 0.35 & 0.40 & 212 \\
Persistence & 0.70 & 0.93  & 0.80 & 172 & 0.30 & 0.49 & 0.37 & 377 & 0.43 & 0.42 & 0.42 & 549 \\
Privilege Escalation & 0.89 & 0.82  & 0.86 & 261 & 0.49 & 0.41  & 0.45 & 462 & 0.65 & 0.34 & 0.44 & 723  \\
Reconnaissance & 0.50 & 1.00  & 0.67 & 2 & 0.41 & 0.42  &  0.41 & 89 & 1.00 & 0.04 & 0.08 & 91 \\
Res. Development & 0.75 & 0.66  & 0.70 & 70 & 0.48 & 0.38  & 0.42 & 205 & 0.85 & 0.17 & 0.28 & 275  \\
 \hline
  \textbf{Samples Average} & \textbf{0.85} & \textbf{0.95}  & \textbf{0.88} & \textbf{4635} & \textbf{0.47} & \textbf{0.65} & \textbf{0.52} & \textbf{6317} & \textbf{0.55} & \textbf{0.54} & \textbf{0.54} & \textbf{10952}  \\
\end{tabular}
\end{scriptsize}
\label{table:encoderdecodercomparison}
\end{center}
\end{table*}

\vspace*{-8pt}

In addition, by comparing Prompt-Only results (Table \ref{table:decoderonly_performance}) and w/o Matched URLs (Table \ref{table:encoderdecodercomparison}), we can see when the LLM is provided with additional `distracting' context (RAG), the F1 score is lower than relying solely on the model's pre-trained knowledge. This indicates that RAG puts high attention to the retrieved context (which is inserted into its prompt). 
This again signifies the importance of retrieving directly relevant
context for d-LLM. Note that one could suggest to fine-tune d-LLMs, but doing so for `large' d-LLMs is computationally demanding, which is not feasible for cybersecurity practitioners as they may not have access to such resources. Moreover, even with fine-tuning, the responses could still be `distracting' if the training data is not directly relevant to a specific attack procedure. 

Our experiment design so far aims to have LLMs produce accurate prediction of ATT\&CK tactics. This puts stress on LLMs to look for key phrases to produce an optimal response. This raises the question: Will one achieve better precision and recall if we remove the specifics, namely the ATT\&CK tactics, as part of the prompt? We further explore this hypothesis next.

%


\vspace*{-16pt}

\subsection{Analysis of Specific Cases for Decoder-only w/ RAG}
\label{sec:challenges}

To assess how d-LLM with RAG may respond differently, we design a similar yet generic prompt (i.e., without asking specifically for ATT\&CK tactics) as follows:

\begin{promptbox}

{\it You are a cybersecurity expert. Consider the relevant context provided below and answer the question.

Relevant Context: \{context\}

Question: Knowing that $<<${procedure}$>>$, what will a cyber adversary achieve with this technique?
}
\end{promptbox}

We extracted 20 procedure examples for this study, recognizing the labor-intensive nature and the specialized expertise required to evaluate the responses accurately. We first examine one of the specific procedures, shown below along with its tactic, the RAG retrieved procedures, and part of the retrieved context.

\begin{mybox}[title=Example-MuddyWater]
\footnotesize
\textit{\textbf{\small Procedure in question (tactic: Credential Access):}} 

\textit{\small MuddyWater has performed credential dumping with Mimikatz and procdump64.exe} 

\vspace{2pt}

\textit{\textbf{\small Top-3 similar procedures:}}

1. \textit{APT39 has used Mimikatz, Windows Credential Editor and ProcDump to dump credentials.}

2. \textit{APT32 used Mimikatz and customized versions of Windows Credential Dumper to harvest credentials.}

3. \textit{Okrum was seen using MimikatzLite to perform credential dumping.}

\vspace{2pt}

\textit{\textbf{\small Retrieved context:}}

\textit{...Mimikatz performs credential dumping to obtain account information useful in gaining access to ...}

\textit{\textbf{\small Response from the Model:}}

`Credential Access', `Defense Evasion', `Discovery', and `Lateral Movement'
\end{mybox}

\noindent 
The RAG process for this procedure retrieved 3 URLs, all of which described a technique/sub-technique of `Credential Access'. This matches to the ground-truth tactic (according to MITRE ATT\&CK). Using the original prompt, the d-LLM (GPT-3.5) predicted: `Credential Access', `Defense Evasion', `Discovery', and `Lateral Movement'. Such a scenario exemplifies a situation where RAG is good at `Recall' (i.e., predicting the right tactic) but not `Precision'.
Meanwhile, the generic prompt produces the following:


\begin{mybox}[title=Example-MuddyWater w/ generic prompt]

\textit{\textbf{Response from the Model:}}

\textit{A cyber adversary using the credential dumping technique with Mimikatz and procdump64.exe, as seen with MuddyWater, would be able to obtain account and password information from the compromised system. This information can then be used to gain unauthorized access to additional systems and enterprise network resources, allowing the adversary to move laterally within the network and potentially escalate their privileges. Additionally, the stolen credentials can be used for various malicious activities, including data theft, sabotage, or further compromise of the network.
}

\end{mybox}

\noindent We manually evaluated the above response and assessed that the model predicts `Credential Access' and `Defense Evasion' with potential follow-up adversary actions of `Lateral Movement' and `Privilege Escalation'. Meanwhile, the original prompt almost always gives specific tactics without elaborated sentences. For this example, we believe the generic prompt improves the `precision' of the model because `Credential Access' and `Defense Evasion' are the only immediate outcomes achieved by MuddyWater using Mimikatz and procdump64.exe. Even if we compare all four tactics associated with the generic vs. original prompts, we still think that `Privilege Escalation' makes more sense than `Discovery'.




We repeat this manual analysis for 20 procedures, 10 of which have RAG produce a match to URL and the other 10 do not (recall Table \ref{table:encoderdecodercomparison}). 
Our evaluation for the 10 matched cases suggests that the new generic prompt gives better precision and recall than the original prompt does. For the 10 no-matched cases, the performance seems to remain comparable between the two prompts, which underscores the importance of retrieving directly relevant information for the d-LLM to perform well. In all 20 cases, similar to the MuddyWater case, the generic prompt produces more elaborated responses, which we believe are more valuable to security analysts. 
This preliminary finding seems to suggest the use of d-LLMs with RAG will be better off without too specific of instructions (i.e., asking specifically for ATT\&CK tactics). This is counter-intuitive in that providing more generic prompts could actually lead to better precision. One could hypothesize that this is due to the generative power of d-LLMs, which helps derive more precise final answers. Further systematic and large-scale study is needed to verify this hypothesis. This will be a daunting task that requires labor-intensive cybersecurity expertise.

\section{Concluding Remarks}
\label{sec:conclusion}

We conducted a comparative analysis of the performance in SFT of smaller e-LLMs against RAG-enhanced larger d-LLMs. This comparison focuses on determining which model type is more effective in interpreting cyberattack procedure descriptions. We observed a significant improvement by d-LLMs when supplemented with appropriate RAG inputs. 
Our research examined variants of RAG in refining TTP descriptions, and revealed RAG could be limited when it fail to retrieve the most relevant information. Furthermore, we found that current d-LLMs, while exhibiting high recall, often lack precision. This observation underscores the need for new advancements to boost d-LLM's precision without compromising recall. We further experimented with a more generic prompt, and show that d-LLM with RAG in this case not only elaborated better but also offered higher precision. This last observation is via a small-scale study, due to the needs for extensive expertise and manual assessment, but sheds lights for future research. Practical cyber-ops will benefit from elaborated LLM responses with meaningful contextual reasoning. We envision a computationally efficient, fine-tuned d-LLM with RAG to offer such a capability with generic prompts.

\bibliography{colm2024_conference}
\bibliographystyle{colm2024_conference}

\end{document}